\documentclass[aps,twocolumn,showpacs,superscriptaddress,prl]{revtex4}
\usepackage{graphicx}
\usepackage{amssymb}
\usepackage{amsmath}

\newcommand{\mf}{3\lambda/4}
\newcommand{\mz}{\lambda/4}
\newcommand{\K}{\mathrm K}
\newcommand{\mK}{\mathrm{mK}}
\newcommand{\mm}{\mathrm{mm}}
\newcommand{\nm}{\mathrm{nm}}
\newcommand{\um}{\mu\mathrm{m}}
\newcommand{\kb}{k_{\mathrm{B}}}
\newcommand{\Bp}{B_{\|}}
\newcommand{\gens}{g_{\mathrm{ens}}}
\newcommand{\KHz}{\mathrm{KHz}}
\newcommand{\MHz}{\mathrm{MHz}}
\newcommand{\Hz}{\mathrm{Hz}}
\newcommand{\GHz}{\mathrm{GHz}}
\newcommand{\mT}{\mathrm{mT}}
\newcommand{\dBm}{\mathrm{dBm}}

\newcommand{\s}{\mathrm{s}}
\newcommand{\millis}{\mathrm{ms}}

\newcommand{\Sto}{S_{21}}
\newcommand{\fc}{\omega_{\mathrm{c}}}

\newcommand{\fp}{\omega_{\mathrm{p}}}
\newcommand{\mi}{m_I}

\newcommand{\ki}{\kappa_{\mathrm{i}}}
\newcommand{\ke}{\kappa_{\mathrm{e}}}

\newcommand{\thirteenC}{^{13}\mathrm{C}}

\newcommand{\mo}{m_{0}}
\newcommand{\Tc}{T_{\mathrm{c}}}

\newcommand{\Gammao}{\Gamma_{\mathrm{o}}}

\newcommand{\Omegam}{\Omega_{\mathrm{0}}}

\newcommand{\Nphot}{N_{\mathrm {phot}}}
\newcommand{\Pp}{P_{\mathrm {p}}}
\newcommand{\YSiO}{\mathrm{Y}_2\mathrm{SiO}_5}

\begin{document}

\title{Probing dynamics of an electron-spin ensemble via a superconducting resonator}

\author{V. Ranjan\footnote[1]{These authors contributed equally to this work.}}
\author{G. de Lange\footnotemark[1]}
\author{R. Schutjens}
\affiliation{Kavli Institute of Nanoscience, Delft University of Technology, P.O. Box 5046,
2600 GA Delft, The Netherlands}
\author{T. Debelhoir}
\affiliation{ICFP, D\'epartement de Physique de l'ENS, 24 rue Lhomond, 75005 Paris, France}
\author{J. P. Groen}
\author{D.~Szombati}
\author{D. J. Thoen}
\author{T. M. Klapwijk}
\author{R. Hanson}
\author{L. DiCarlo}
\affiliation{Kavli Institute of Nanoscience, Delft University of Technology, P.O. Box 5046,
2600 GA Delft, The Netherlands}
\date{\today}

\begin{abstract}
We study spin relaxation and diffusion in an electron-spin ensemble of nitrogen
impurities in diamond at low temperature ($0.25$-$1.2~\K$) and polarizing
magnetic field ($80$-$300~\mT$). Measurements exploit mode- and temperature-
dependent coupling of hyperfine-split sub-ensembles to the resonator.
Temperature-independent spin linewidth and relaxation time suggest that spin
diffusion limits spin relaxation. Depolarization of one sub-ensemble by resonant
pumping of another indicates fast cross-relaxation compared to spin diffusion,
with implications on use of sub-ensembles as independent quantum memories.
\end{abstract}


\maketitle

The study of spin ensembles coupled to superconducting integrated circuits is of both technological and fundamental interest. An eventual quantum computer may involve a hybrid architecture~\cite{Wesenberg2009,Marcos2010,Zhu2011,Kubo2011} combining superconducting qubits for processing of information,
solid-state spins for storage, and superconducting resonators for interconversion. Additionally, superconducting resonators allow the study of spin ensembles at low temperatures with ultra-low excitation powers and high spectral
resolution~\cite{Malissa2012,Kubo2012}. While one spin couples to one microwave photon with strength $g/2\pi \sim 10~\Hz$, an ensemble of $N$ spins collectively couples with $\gens=g\sqrt{N}$~\cite{Imamoglu2009,Amsuss2011}, reaching the strong-coupling regime $\gens > \kappa, \gamma$ at $N\gtrsim10^{12}$~\cite{Kubo2010, Schuster2010,Amsuss2011}, where  $\kappa$ and $\gamma$ are the circuit damping and spin dephasing rates, respectively.

Among the solid-state spin ensembles under consideration, nitrogen defects in diamond (P1 centers)~\cite{Loubser1978} are excellent candidates for quantum information processing. Diamond samples can be synthesized with P1 centers as only paramagnetic impurities. Additionally, samples with spin densities ranging from highly dense ($>200$~ppm)
to very dilute ($< 5$~ppb) are commercially available, allowing the tailoring of
spin linewidth ($\gamma \propto N$ \cite{vanWyk1997}) and collective strength
($\gens \propto \sqrt{N}$). In contrast to nitrogen-vacancy centers in
diamond~\cite{Dutt2007} and rare-earth ions in
$\YSiO$~\cite{Bushev2011,Staudt2012}, P1 centers are optically inactive, making
a coupled microwave resonator an ideal probe for their study. However, the
magnetic fields $\gtrsim 100~\mT$ needed to polarize the ensemble at the
few-$\GHz$ transition frequencies of circuits~\cite{Clarke2008} must not
compromise superconductivity. The freezing of all spin dynamics in a high-purity
P1 ensemble by the field would allow quenching  spin
decoherence~\cite{Takahashi2008} through dynamical decoupling~\cite{Lange2012},
realizing a useful quantum memory.

\begin{figure}[b!] \includegraphics[width=\columnwidth]{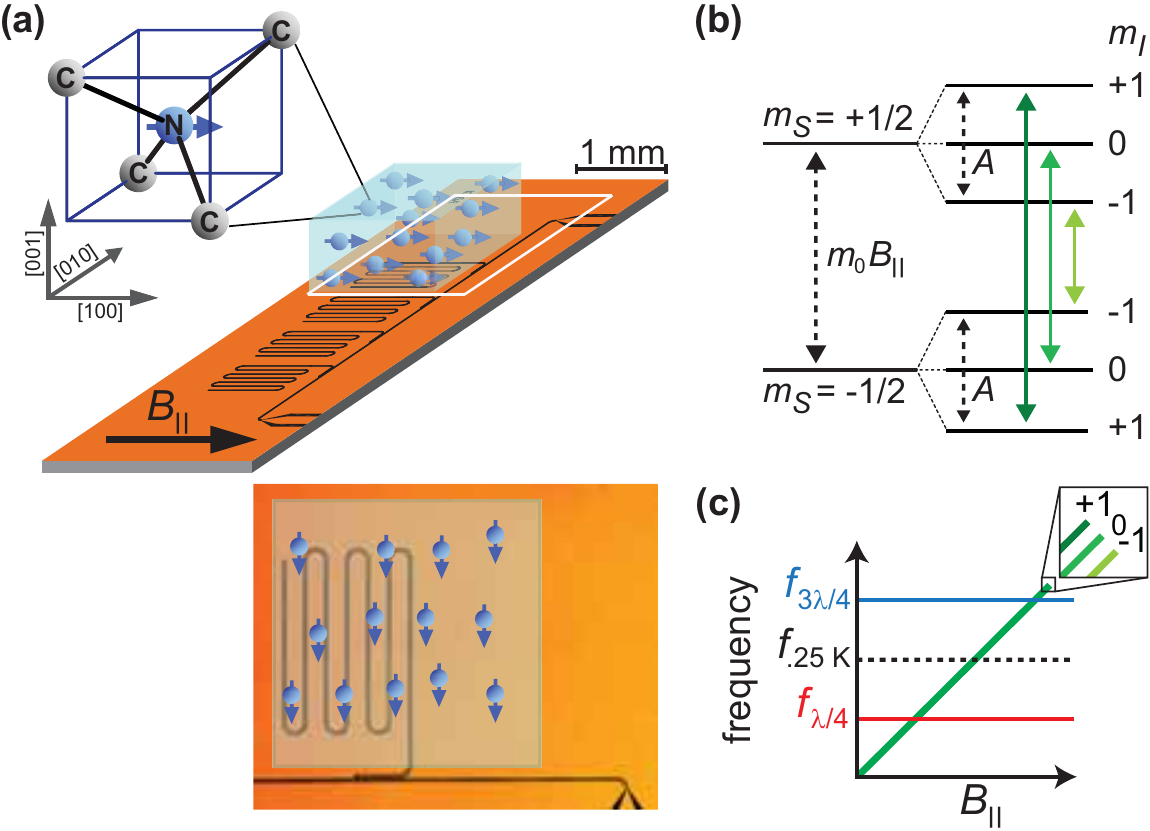}
\caption{(color online). (a) Schematic of the hybrid resonator-spin system. A  single-crystal diamond piece ($1.7~\mm\times1.7~\mm\times1.1~\mm$, type-Ib Sumicrystal, $\sim 200$~ppm N content) is placed on top of one of four CPW resonators capacitively coupled to a common feedline.  The resonators are patterned on a NbTiN film ($70~\nm$ thick, critical temperature $\Tc=12.5~\K$) on sapphire (C-plane, $430~\um$ thick). An external magnetic field $\Bp$ is applied parallel to the film, along the diamond [100] direction. (b) Hyperfine interaction $A\approx 94~\MHz$  with the N host nuclear spin splits each electron-spin level into a triplet. Only electron-spin transitions that preserve nuclear spin (solid arrows) are allowed. (c) $\Bp$ tunes the electron-spin energy levels through resonance with the $\mz$ or $\mf$ modes of the resonator. The dashed line represents the thermal energy $\kb T/h\approx5~\GHz$ at $T=0.25~\K$.}
\label{fig:1sample}
\end{figure}

Here, we investigate the dynamics of a P1 electron-spin ensemble
probed by controlled coupling to  two modes of a coplanar waveguide (CPW)
resonator. The resonator is patterned on a NbTiN film~\cite{Barends2010}
withstanding applied magnetic fields beyond $300~\mT$.  Three hyperfine-split
spin sub-ensembles are clearly resolved over the temperature range
$0.25$-$1.2~\K$.  The collective coupling of each sub-ensemble to the resonator
reaches $\gens=17~\MHz$ at $250~\mK$, in accordance with field- and
temperature-controlled spin polarization, and extrapolates to $\gens=23~\MHz$ at
full polarization. Using a pump-probe technique in the dispersive regime, we
measure spin linewidth and relaxation time. The observed temperature
independence indicates that internal spin equilibration is dominated by spin
diffusion across the mode volume~\cite{Bloembergen1959} rather than spin-lattice
relaxation~\cite{Reynhardt1998}. Finally, as an initial test of the possible use
of sub-ensembles as independent memories, we measure the steady-state
depolarization of a sub-ensemble by resonant pumping of another. The pump-power
dependence observed indicates fast cross-relaxation compared to spin relaxation
in the mode volume, calling for follow-up experiments probing the millisecond
scale.

Our hybrid system, shown schematically in Fig.~\ref{fig:1sample}, consists of four resonators capacitively coupled to a common feedline and a type-Ib diamond sample placed above one of them. The electron-spin ensemble consists of unpaired electrons (spin-1/2) at substitutional nitrogen impurities [Fig.~\ref{fig:1sample}(a)]. Each electron spin exhibits strong anisotropic hyperfine interaction with the host nucleus (spin-1). The Hamiltonian for one defect is given by $H_{\mathrm N} = -\mo\vec{B} \cdot \vec{S} + h  \vec{S} \cdot A \cdot \vec{I}$, with $\vec{S}$ and $\vec{I}$ the spin operators for the electron and nitrogen nucleus, respectively, $\mo/h = 28.0~\MHz/\mT$,  $h$ Planck's constant, and  $A=\mathrm{diag}(81.33, 81.33, 114.03)~\MHz$ the hyperfine interaction tensor~\cite{Cook1966} [third (first, second) index parallel (normal) to the Jahn-Teller axis]. Low-energy terms only involving $\vec{I}$  have been left out. We tune the electron-spin transitions with a magnetic field ($\Bp$) applied along the diamond $[100]$ direction. Because all
N-C bonds have $\langle 111 \rangle$ orientation and make the same angle with $\Bp$, the hyperfine interaction is the same  for all impurities, creating three hyperfine-split electron-spin transitions~\cite{Loubser1978}.

\begin{figure}[t] \includegraphics[width=\columnwidth]{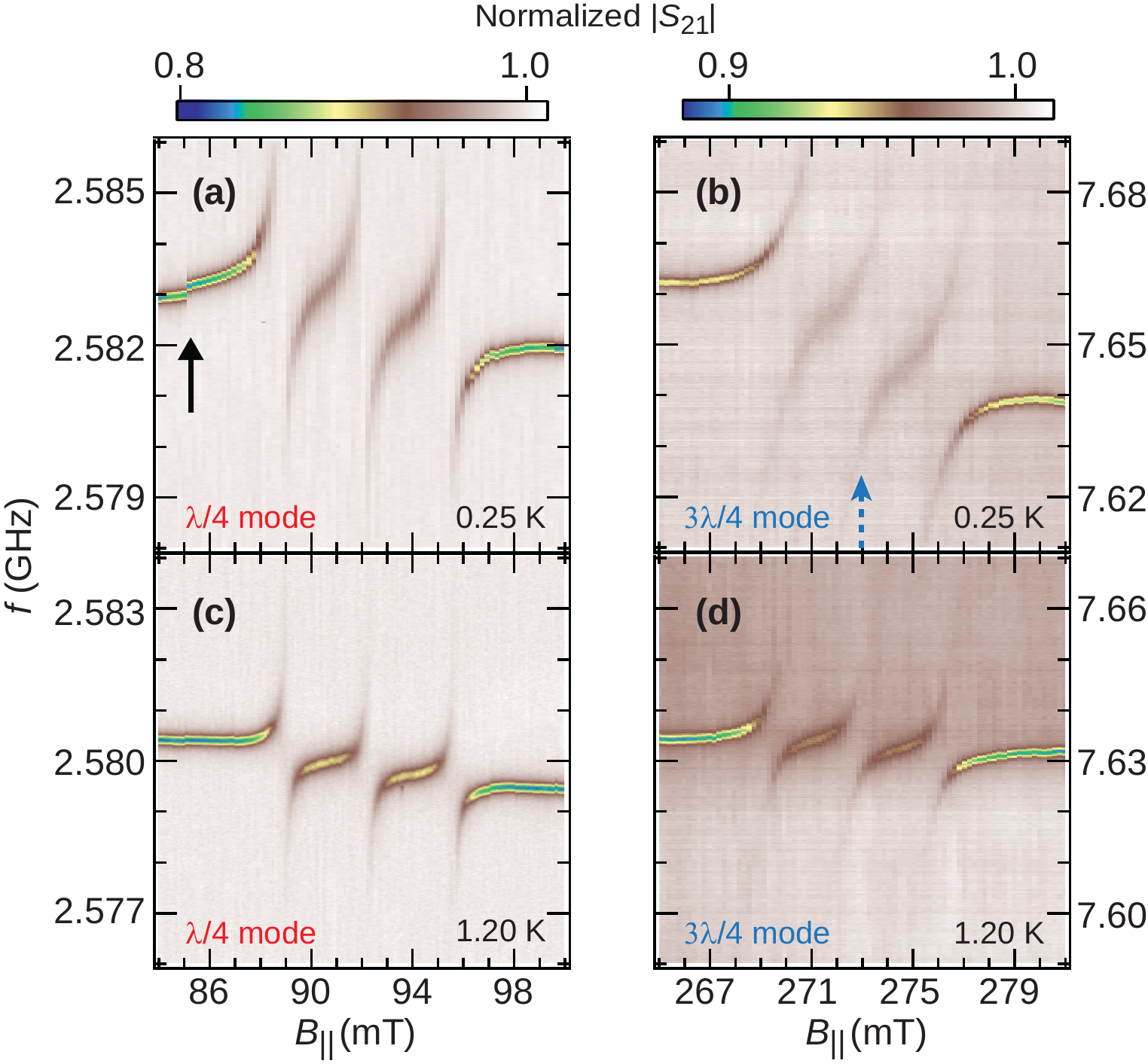}
\caption{(color online). Transmission spectroscopy. Image plots of normalized feedline transmission as a function of $\Bp$
and frequency near the $\mz$ and $\mf$ mode resonances at $T=0.25~\K$ (a,b) and $1.2~\K$ (c,d). Each plot reveals three
avoided crossings, corresponding to allowed hyperfine-split electron-spin
transitions. Note that the frequency span in (b,d)  is 10 times larger
than in (a,c). The arrow in (a) points to a flux jump shifting the resonator frequency. All other
image plots shown are corrected for these rare events.}
\label{fig:2AC} \end{figure}

\begin{figure}[t] \includegraphics[width=\columnwidth]{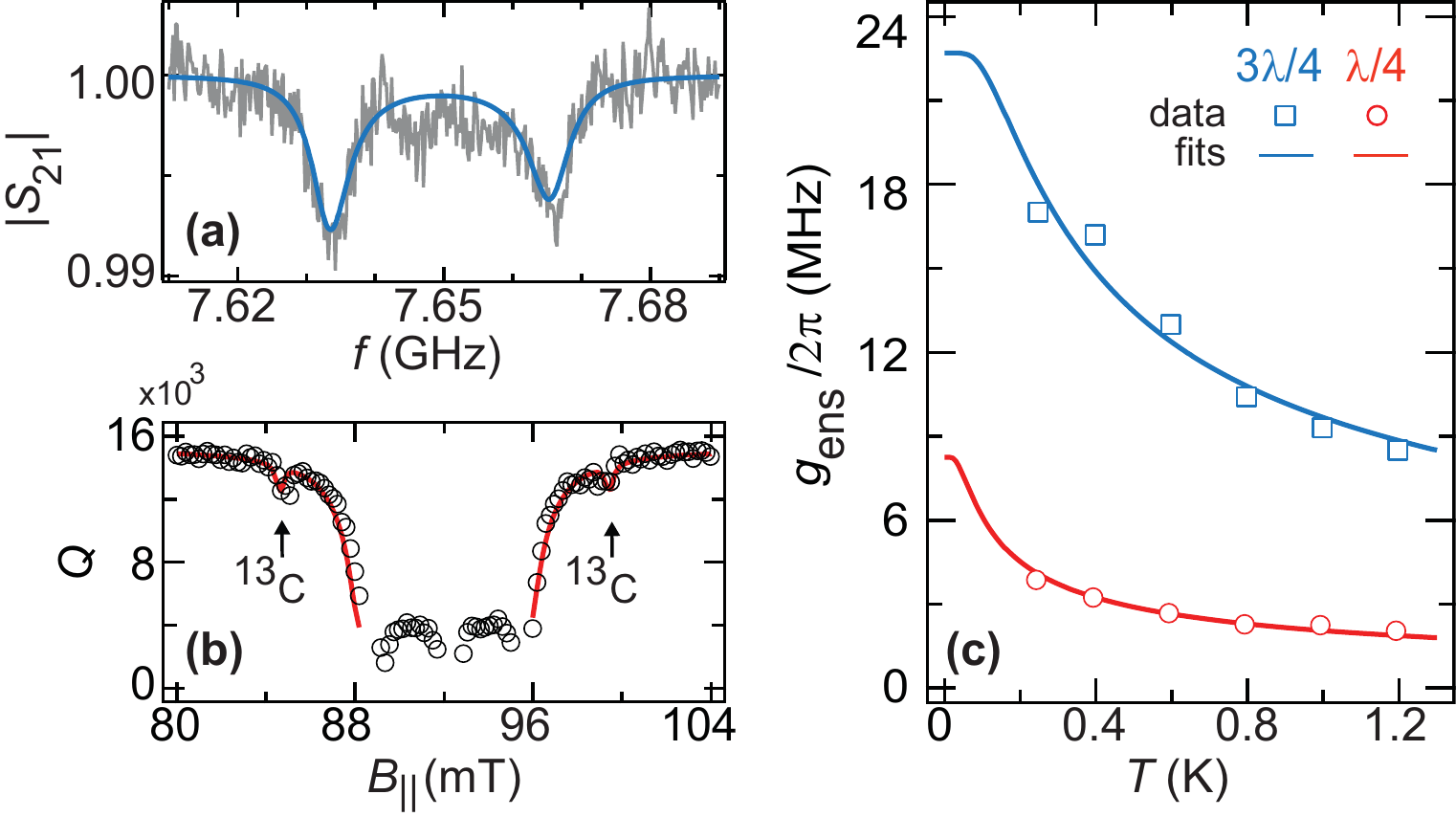}
\caption{(color online). Determination of the collective coupling strength $\gens$.
(a) A vertical cut of Fig.~2(b) at $\Bp=272.8~\mT$ (dashed arrow) shows Rabi-split transmission dips.
The best fit to Eq.~(1) gives $\gens =17.0~\MHz$.  (b) Measured loaded quality factor
of the $\mz$ mode as a function of $\Bp$ at $T=0.25~\K$.  The best fit of Eq.~(1) away from the
avoided crossings gives $\gens=3.9~\MHz$. Arrows point to satellites resulting from the hyperfine coupling of the electron spin to the nuclear spin of $\thirteenC$ atoms adjacent to some P1 centers~(see \cite{SOM} for further discussion). (c) Best-fit $\gens$ to the $\mz$ (circles) and $\mf$ (squares) modes as a function of temperature. Solid curves are the best fits of Eq.~(2). Error bars are smaller than the symbol size.
\label{fig:3gvsT}}
\end{figure}

Measurements~\cite{SOM} of the feedline transmission $|\Sto|(f,\Bp)$ near the fundamental ($\mz$) and the second-harmonic ($\mf$) modes at $T=0.25~\K$ and $1.2~\K$ clearly show three avoided crossings~\cite{Schuster2010}, as expected for coherent coupling~\cite{Abe2011} (Fig.~\ref{fig:2AC}). The coupling strength of each hyperfine transition to the $\mf$ mode is evidently stronger than to the $\mz$ mode, and decreases for both modes with increasing temperature. The hybridized dips observed when spin sub-ensembles are resonant with the $\mf$ mode [Fig.~\ref{fig:3gvsT}(a)] support strong coupling $(2\gens > \gamma, \kappa)$. The absence of double dips on resonance with the $\mz$ mode  indicate  $2\gens <\gamma$.

We extract $\gens$ using the model presented in Ref.~\onlinecite{Schuster2010}, treating the spin sub-ensembles as separate harmonic oscillators coupled to the resonator, but not to each other:
\begin{equation}
    S_{21}(\omega) = 1 + \frac{\ke/2}{i\Delta_{\mathrm c} - \left(
    \ki + \ke \right)/2
    +\sum_n\frac{\gens^2}{i\left(\Delta_{n}\right) -\gamma/2}}.
        \label{eq:S21}
\end{equation}
Here, $\Delta_{\mathrm c} = \omega - \fc$ is the frequency detuning between the probe and bare resonator mode, $\ki$ and $\ke$ are the resonator intrinsic and extrinsic dissipation rates, $\Delta_{n}=\omega-\omega_{\mi=n}$ is the probe detuning from the $\mi=n$ hyperfine transition and $\gamma$ is the transition linewidth (assumed independent of $\mi$). As shown in Figs.~\ref{fig:3gvsT}(a) and \ref{fig:3gvsT}(b), fitting the double-dip spectrum for the $\mf$ mode and the quality factors $(Q)$ for $\mz$ mode at $0.25~\K$ using Eq.~(\ref{eq:S21}) yields collective coupling strengths $\gens/2\pi =17.0\pm 0.4~\MHz$ and $3.9 \pm 0.2~\MHz$, respectively~\cite{SOM}.

To investigate the temperature dependence of the collective coupling strength to each mode, we measure transmission spectra at several temperatures in the range $0.25$-$1.2~\K$ and perform the same analysis as above~\cite{Bushev2011,Sandner2012,Staudt2012}. The results are shown in Fig.~\ref{fig:3gvsT}(c) together with the best
fits to
\begin{equation} \gens(T) = \gens(0) \sqrt{P(\Bp,T)}\ , \ \
\label{g}
\end{equation}
where $\gens(0)$ is the zero-temperature coupling strength, $P(\Bp,T)=\tanh\!\left(\mo \Bp/2\kb T\right)$ the spin polarization in thermal equilibrium, and $\kb$ the Boltzmann constant. Two factors combine to make $\gens(T)$ higher for the $\mf$ mode. First, $P$ increases monotonically with the Zeeman energy
$\mo\Bp$. Second, the bare spin-coupling strength $g$ increases as $\sqrt{\fc}$ owing to a larger vacuum magnetic field strength. The ratio 2.7 between the best-fit
$\gens(0)/2\pi$ values for the $\mf$ and $\mz$ modes ($22.7 \pm 0.6$ and $8.3\pm 0.2~\MHz$, respectively) differs from  the expected $\sqrt{3}$. This discrepancy may be due to inhomogeneous distribution of P1 centers in the mode volume~\cite{Burns1990} (see further below).

\begin{figure}[t]
\includegraphics[width=\columnwidth]{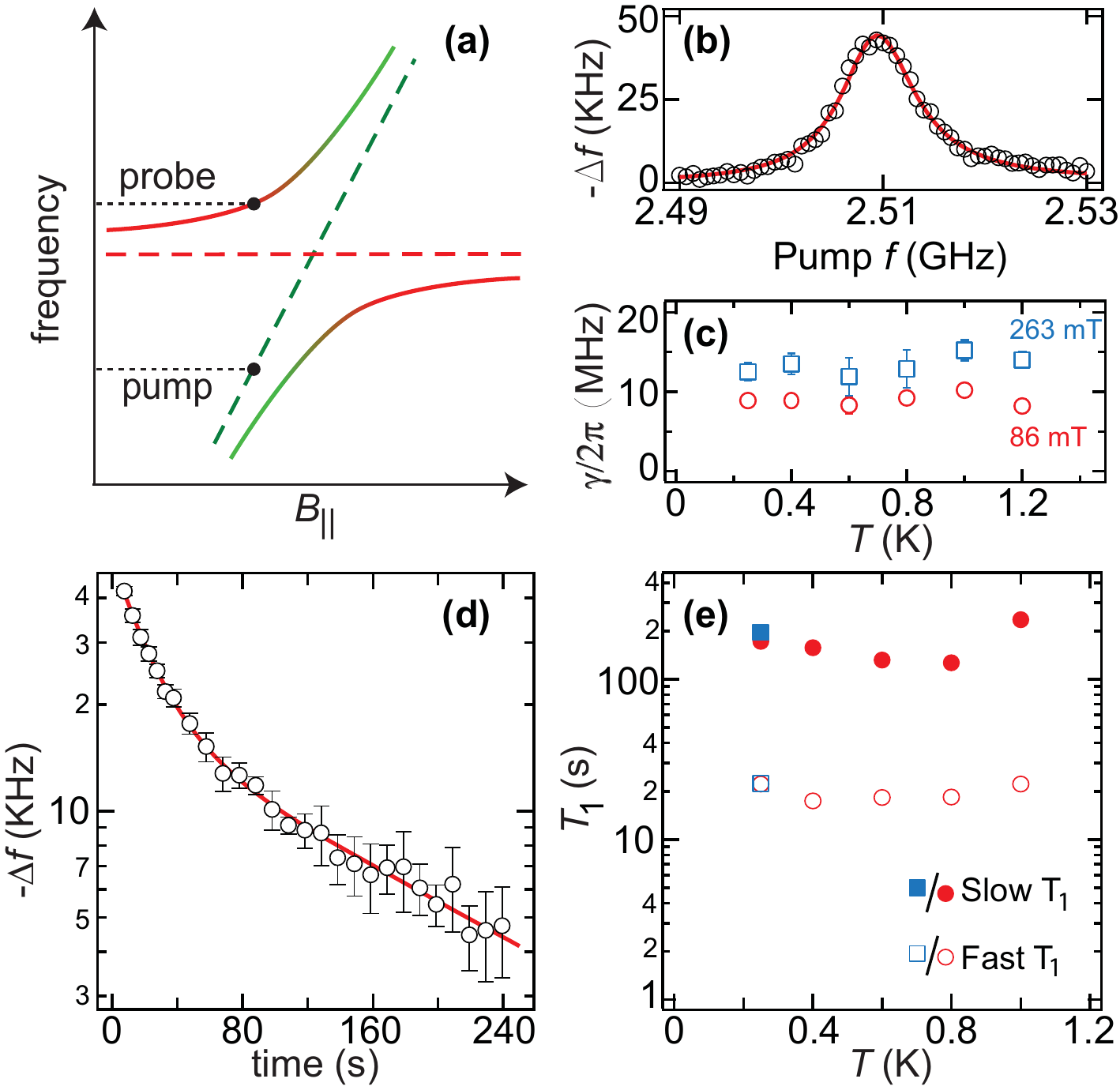}
\caption{(color online) Measurement of the spin linewidth and relaxation
times using dispersive spin-resonator interactions. (a) Scheme (not to scale)
for (b) the measurement of the spin linewidth ($T=0.25~\K$, $\Bp=86~\mT$)
by probing the frequency shift of the $\mz$ mode after applying a pump pulse ($0.4~\s$
duration, $-50~\dBm$ incident power) through resonance with the
$m_{I}=+1$ transition ($60~\s$ wait between successive measurements)~\cite{SOM}.
A similar measurement of $\gamma$ at $\Bp=263~\mT$ is obtained using the
$\mf$ mode. (c) $\gamma$ at $\Bp=86~\mT$ (circles) and $\Bp=263~\mT$
(squares) as a function of temperature.  (d) Measurement of the spin
relaxation time $T_1$ by probing the resonator shift as a function of time after
the pump pulse is switched off. A bi-exponential decay is observed.  (e) Temperature
dependence of the two time constants, extracted by probing with the $\mz$ (circles) and $\mf$ (squares) modes.
Error bars, unless shown, are smaller than the symbol size.}
\label{fig:5LWT1}
\end{figure}

Having characterized coherent coupling in the hybrid system, we now turn to using the resonator as a probe of spin dynamics and equilibration. We first measure linewidth $\gamma$ of the $\mi=+1$ transition in the dispersive regime~\cite{Amsuss2011}, with $\sim70~\MHz \gg \gens$ detuning between the $\mz$ mode and $\mi=+1$ transition. We extract $\gamma$ by inferring~\cite{SOM} the frequency shift ($\Delta f$) of the $\mz$ mode immediately following a pump pulse whose frequency $\fp$ is stepped through resonance with the $\mi=+1$ transition [Fig.~\ref{fig:5LWT1}(b)]. The pump pulse slightly decreases the polarization of the ensemble, red-shifting the resonator. We fit a Lorentzian lineshape to $|\Delta f|$, finding a full-width-at-half-maximum $\gamma/2\pi = 9.0 \pm 0.3~\MHz$. A similar dispersive measurement using the $\mf$ mode at $\Bp=263~\mT$ gives $\gamma/2\pi=12.0\pm0.7~\MHz$. We find these values to be temperature independent in the range $0.25$-$1.2~\K$ [Fig.~\ref{fig:5LWT1}(c)], indicating that $\gamma$ is limited by dipolar interactions and field inhomogeneity. We attribute the $\gamma$ increase with $\Bp$ to the latter.

The  spin relaxation time is measured by applying a pump pulse resonant with the $\mi=+1$ transition and monitoring the frequency shift in time as the spin polarization returns to equilibrium. We observe a bi-exponential decay response with time constants $\sim 20~\s$ and $\sim 160~\s$ [Fig.~\ref{fig:5LWT1}(d)]. These constants are independent of temperature in the range $0.25$-$1~\K$ [Fig.~\ref{fig:5LWT1}(e)], suggesting that spin polarization decay is not limited by spin-lattice relaxation~\cite{Reynhardt1998} but spin diffusion instead. Through dipolar flip-flop processes, the depolarization diffuses out of the resonator mode volume, leading to repolarization of the ensemble. The rate for this process depends on the nominal dipolar coupling strength between spins in the ensemble, which itself depends on the spin density~\cite{vanWyk1997}. The two time constants may be explained by two diamond sectors inside the mode volume with electron-spin densities differing by a factor of $\sim 8$~\cite{Sorokin1960,Burns1990,SOM}. This is supported by the fact that relative amplitudes of the two exponentials do not change versus temperature.

\begin{figure}[t]
\includegraphics[width=\columnwidth]{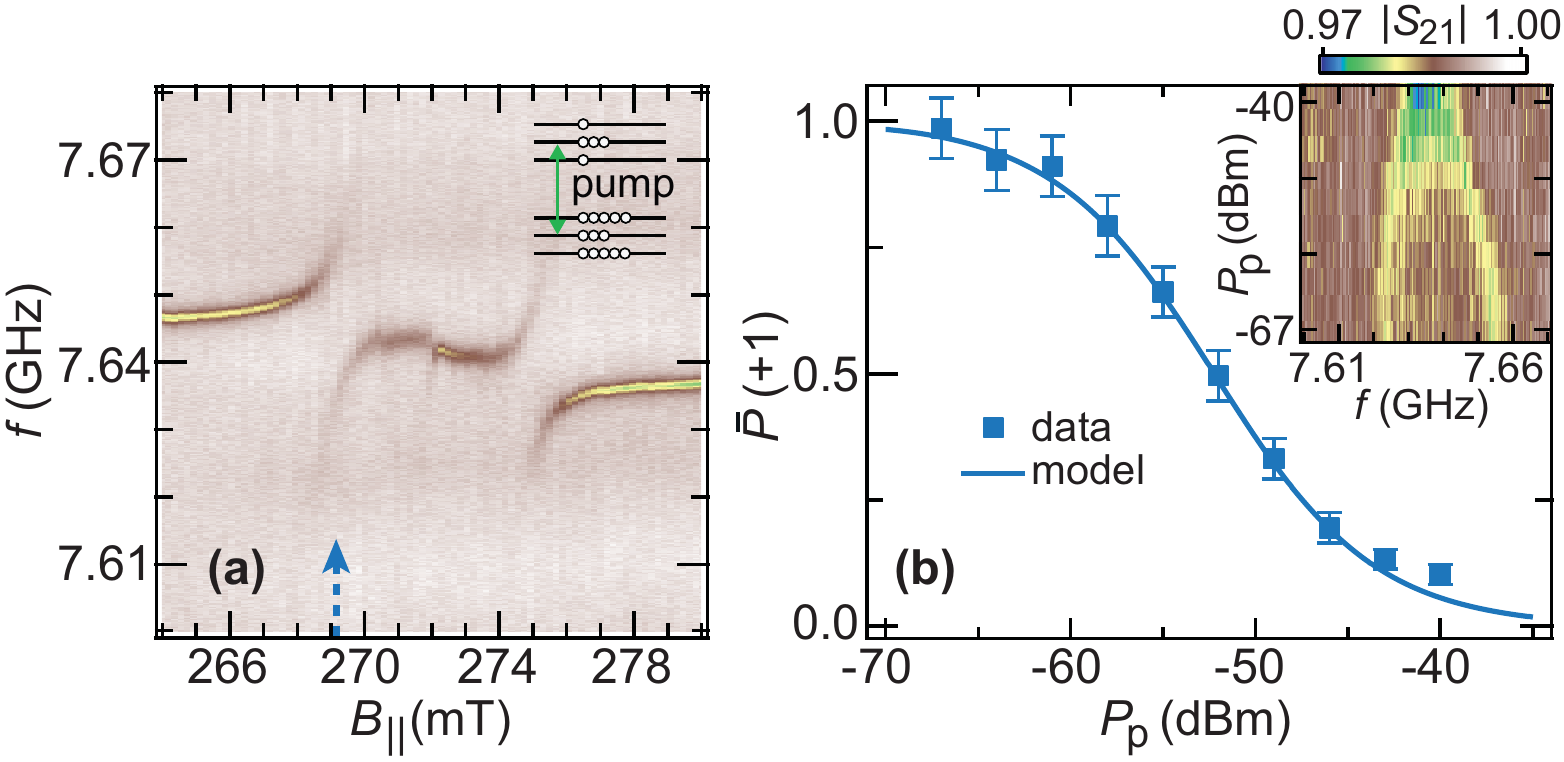}
\caption{(color online). (a) Transmission spectroscopy similar to Fig.~\ref{fig:2AC}(b), with an additional pump pulse resonant with the $\mi=0$ transition (incident pump power $\Pp=-50~\dBm$, $100~\millis$ duration) prior to $|\Sto|$ measurement. A complete disappearance of the $\mi=0$ avoided crossing and a reduction in the coupling strength of the undriven transitions are observed. Color scale is identical to Fig.~\ref{fig:2AC}(b). (b) Inset: vacuum-Rabi-split dips at $\Bp=269.1~\mT$ as a function of $\Pp$. The merging of dips with increasing $\Pp$ indicates cross-relaxation between the sub-ensembles. Main panel: Extracted polarization $\bar P$ (normalized to the value without pump) for the undriven $\mi=+1$ sub-ensemble. The curve corresponds to the steady-state solution of a rate equation modeling fast equilibration between the sub-ensembles compared to $T_1$ (see text and \cite{SOM} for details).}
\label{VAC}
\end{figure}

To investigate spin dynamics across sub-ensembles, we measure how pumping one sub-ensemble can affect the coupling strength of other sub-ensembles to the resonator~\cite{SOM}. As shown in Fig.~\ref{VAC}(a), pumping at $f_{\mi=0}(\Bp)$ completely suppresses the avoided crossing between the $\mi=0$ transition and the resonator~\footnote{This is surprising from a single-spin perspective, because the maximum Rabi driving strength ($f_{\mathrm {Rabi}}= g\sqrt{\Nphot}/2\pi \approx 100~\KHz$ for $\Nphot=10^8$ photons on mode resonance) is significantly smaller than the spin linewidth.}. Remarkably, partial depolarization is evident in the $\mi=\pm1$ sub-ensembles. The coupling strengths of the undriven transitions $(\mi=\pm1)$ to the $\mf$ mode are reduced to $\gens/2\pi = 12.5 \pm 0.5$ and $12.0\pm0.5~\MHz$, respectively. To quantify this steady-state cross-relaxation, we measure the minimum-splitting between the hybridized dips at $\Bp=269.1~\mT$ [arrow in Fig.~\ref{VAC}(a)] as a function of pump power $\Pp$.  As shown  in the inset of Fig.~\ref{VAC}(b), the undriven $\mi=+1$ sub-ensemble depolarizes further with increasing $\Pp$. We can reproduce~\cite{SOM} this power-dependent steady-state cross-depolarization using a rate equation including a spin diffusion rate $\Gammao$ across the mode volume and a cross-relaxation rate $\Gamma$ between sub-ensembles~\cite{Bloembergen1959}. We assume $\Gamma\gg\Gammao$ consistent with previous measurements of cross-relaxation in high density P1-center samples by Sorokin~\textit{et al.}~\cite{Sorokin1960}. Under these assumptions, the steady-state normalized polarization of each sub-ensemble is $\bar P$ = $\Gammao/(\Gammao+\Omegam/3)$, where $\Omegam$ is the pumping rate for the $\mi=0$ transition. Excellent agreement is found with the  model, with only the lever arm between $\Omegam$ and $\Pp$ as free parameter. Using the best-fit lever arm in combination with Fermi's golden rule $\Omegam= 2\pi g^2 \Nphot/\gamma$ and the measured $\Gammao \approx 0.05 ~\s^{-1}$ and $\gamma/2\pi \approx 12~\MHz$, we estimate $g\sim 2.5~\Hz$ \footnote{Note that $\Nphot$ is lower than on mode resonance by the filter factor $(\ki+\ke)^2/(\fc-\fp)^2$.}. Comparing this $g$ to $\gens(T=0)$ suggests $N\sim 10^{14}$ spins in the resonator mode volume.

In conclusion, we have used resonant and dispersive interactions with the two lowest-frequency modes of a NbTiN CPW resonator to probe the dynamics of a P1 electron-spin ensemble in diamond at low temperature and polarizing magnetic field. The observed temperature independence of spin linewidth and relaxation in the range $0.25$-$1~\K$ supports spin out-diffusion as the dominant relaxation mechanism within the resonator mode volume. Resonant pumping of spin sub-ensembles indicates exchange of Zeeman and dipolar energies between sub-ensembles~\cite{Bloembergen1959}. Follow-up experiments will pursue two directions: probing sub-ensemble response to one or more resonant pump pulses on millisecond timescales to shed light on
the cross-relaxation mechanism, and cooling to $15~\mK$ to fully polarize the ensemble~\cite{Takahashi2008} and
extend spin coherence using sub-ensemble-selective dynamical decoupling~\cite{Lange2012}. Ultimately, cross-relaxation and achieved coherence will set the timescale over which sub-ensembles may serve as independent quantum memories.

\begin{acknowledgments}
We thank D. Rist\`e for experimental assistance and D.~I.~Schuster for helpful comments on the manuscript. 
We  acknowledge funding from the Dutch Organization for Fundamental Research on Matter (FOM),
the Netherlands Organization for Scientific Research (NWO, VIDI scheme),
and a Marie Curie Career Integration Grant (L.D.C.).
\end{acknowledgments}

\end{document}

\title{Supplement to ``Probing dynamics of an electron-spin ensemble via a superconducting resonator"}

\author{V. Ranjan\footnote[1]{These authors contributed equally to this work.}}
\author{G. de Lange\footnotemark[1]}
\author{R. Schutjens}
\affiliation{Kavli Institute of Nanoscience, Delft University of Technology, P.O. Box 5046,
2600 GA Delft, The Netherlands}
\author{T. Debelhoir}
\affiliation{ICFP, D\'epartement de Physique de l'ENS, 24 rue Lhomond, 75005 Paris, France}
\author{J. P. Groen}
\author{D.~Szombati}
\author{D. J. Thoen}
\author{T. M. Klapwijk}
\author{R. Hanson}
\author{L. DiCarlo}
\affiliation{Kavli Institute of Nanoscience, Delft University of Technology, P.O. Box 5046,
2600 GA Delft, The Netherlands}
\date{\today}

\maketitle

\section*{MEASUREMENT SETUP}
The device is operated in a $\threeHe$ refrigerator with base temperature $T=0.25~\K$. All measurements involve continuous-wave heterodyne detection of feedline
transmission  $\Sto$ as a function of $\Bp$ and $T$ with $1~\MHz$ intermediate frequency and $80~\millis$ integration. Incident power
on the feedline ranges from $-115$ to $-95~\dBm$ ($-110$ to $-90~ \dBm$) for the $\mz$ ($\mf$) mode, corresponding to $\sim 10^3$ to $10^{5}$ intra-cavity photons on resonance (incident power for one photon on resonance $\approx \hbar \fc (\ki+\ke)^2/2\ke$~\cite{barendsphd}). All other microwave pulses are also applied through the feedline. At $\Bp=0$ the diamond-coupled resonator has $\mz$ and
$\mf$ modes at $2.59$  and $7.67~\GHz$, respectively ($\sim11\%$ lower than the values without the diamond). The reduction in frequency of the resonator after mounting diamond is due to the higher effective dielectric constant of the diamond $(\sim 5.7)$. The corresponding loaded
quality factors are $\Ql=\fc/(\ki+\ke) = 1.8\times 10^{4}$ and $6\times 10^{3}$ ($\sim3$ times lower than the values without the diamond). Increase in intrinsic losses could come from the surface of the diamond and from the vacuum grease used to glue the diamond to the resonator. Increasing $\Bp$ to $300~\mT$ further decreases $\Ql$ by $\sim 15\%$. To reduce
microwave losses in the superconducting film resulting from trapped flux, the system is frequently reset by warming above the film $\Tc$ and cooling back down with $\Bp=0$.

\section*{AVOIDED CROSSINGS}
Transmission spectroscopy measurements in Fig.~2 reveal two regimes of spin-resonator coupling.
Rabi-split dips are clearly resolved [Fig.~3(a)] when the $\mi=0$ transition is resonant with the $\mf$ mode, evidencing strong coupling $(2\gens> \gamma,\kappa)$.
To extract $\gens$ in this case, we fit Eq.~(1) to $|\Sto|(f)$~\cite{Schuster2010}. For this fit, we fix $\ki$, $\ke$ and $\gamma$ to independently-measured values: $\ki/2\pi=1.4\pm 0.1~\MHz$  and $\ke/2\pi=120\pm10~\kHz$ are extracted on both sides of the avoided crossings, and $\gamma/2\pi=12.0\pm0.7~\MHz$ is obtained from dispersive pump-probe linewidth measurements (Fig.~4).

The $\mz$ mode is weakly coupled to the resonator ($2\gens<\gamma$) and Rabi-split dips are not resolved on resonance. The extraction of a coupling strength to this mode involves several steps. First, we extract the field-dependent loaded quality factor $\Ql$ by fitting a lineshape~\cite{barendsphd}
\[
\Sto(\omega)=\frac{\Sto^{\mathrm{min}}+i2\Ql(\omega-\fr)/\fr}{1+i2\Ql(\omega-\fr)/\fr}
\]
to the measured $|\Sto|(\omega)$ at each magnetic field. Figure~3(b) shows $\Ql(\Bp)$ near the avoided crossings for the $\mz$ mode at $T=0.25~\K$.
The decrease in $\Ql$ when the spin transitions are tuned into resonance with the resonator is caused by  absorption of microwave energy by the spin ensemble.
An analytic approximation of $\Ql=\fr/\Delta \fr$ valid away from the avoided crossings can be obtained from Eq.~(1) and noting that on resonance, where $\abs{S_{21}}$ is minimum, ${\rm Im}(S_{21}) \approx 0$.
Here,

\[
\Delta \fr = (\ki+\ke)+\sum_n \gens^2 \frac{\gamma/2}{\left(\Delta_{\mathrm{n}}^2+\gamma^2/4\right)}
\]
and $\Delta_{\mathrm n} = \omega_{\rm r}-\omega_{\mi=n}$. The fixed
values, $\gamma/2\pi =9.0\pm0.3~\MHz$, $\ki/2\pi=140 \pm 10~\kHz$ and $\ke/2\pi=32 \pm 1~\kHz$, are obtained by the same methods as described for the $\mf$ mode.

The extraction of $\gens$ requires accounting for all ensembles in the last term of $\Delta{\fr}$. We also observe two small dips $\sim 4.1~\mT$ away from the avoided crossings with the $\mi=\pm 1$ transitions [indicated by arrows in Fig.~3(b)]. These dips are consistent with coupling to the nuclear spin ($I =
1/2$) of a $\thirteenC$  (1.1\% natural abundance) at the nearest-neighbor carbon site present on the Jahn-Teller axis ($A_{\thirteenC,\rm a} =
\mathrm{diag}(141.8,141.8,340.8)~\MHz$~\cite{Cox1994}). These give rise to six sub-ensembles with coupling strengths $g^{\rm a}_{\rm ens,\thirteenC} \approx \gens \sqrt{0.011/2} $. The factor $1/2$ comes from the fact that $\thirteenC$ has nuclear spin of $I=1/2$ in contrast to $^{14} {\mathrm N} ~(I=1)$
Additionally, there are six more $\thirteenC$ transitions with $A_{\thirteenC,\rm b} =\mathrm{diag}(32.1,32.1,41.0)~\MHz$~\cite{Cox1994}, which arise from $\thirteenC$ located at the three remaining nearest-neighbor sites. These are three times more abundant with $g^{\rm b}_{\rm ens,\thirteenC} \approx  \sqrt{3} g^{\rm a}_{\rm ens,\thirteenC}$. Although they are not visible as separate dips in the data, they do modify $Q(\Bp)$. We therefore include all of these sub-ensembles in our modeling.

One can in principle also find an analytical expression for the dispersive shift of the resonance with respect to $\fc$ and use it to extract $\gens$ for the $\mz$ mode. In our experiments,
however, this is problematic due to sudden changes in $\fc$ [indicated by the arrow in Fig.~2(a)] caused by flux jumps in the NbTiN film. The absorption, however, only depends on $\fr$, which is extracted with high precision from the transmission lineshape.

\section*{DISPERSIVE MEASUREMENTS}
When performing the dispersive measurements shown in Fig.~4, we measure the change in $|\Sto|$ at fixed frequency $f=\left(\fr-\Delta\fr/2\right)/2\pi$. Immediately before the pumping step, we measure the resonator lineshape and calibrate the slope of $|\Sto|$ at $f$. We use this calibration to infer the resonator shift $\Delta f$. To ensure linearity within $10\%$, the pump pulse power and duration are chosen as to keep the maximum $|\Delta f|\lesssim \Delta\fr/8 \pi$.

\begin{figure}[t]
\includegraphics[width=0.825\columnwidth]{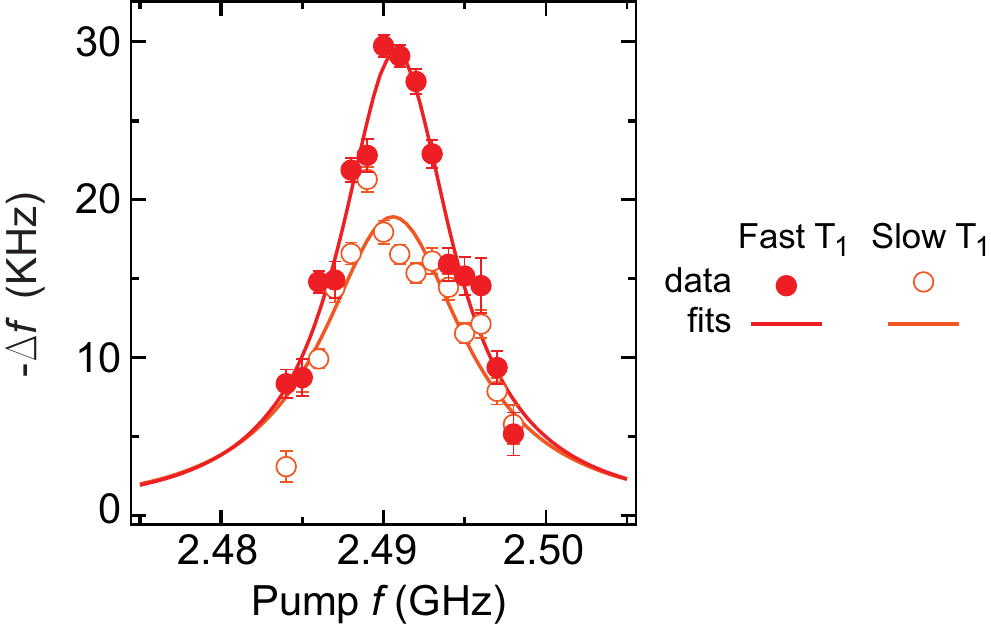}
\caption{Best-fit amplitudes of the two exponential terms observed in the relaxation measurement as function of pump frequency. Lorentzian fits have widths $\gamma_1/2\pi=8.5 \pm 0.5~ \MHz$ (solid circles) and $\gamma_2/2\pi=10.7 \pm 1.5 ~\MHz$ (open circles) for fast and slow decays, respectively. Within error bars, these widths are consistent with the spin linewidth measurements in Fig.~4(b-c).}
\label{T1}
\end{figure}

Measuring the relaxation of pumped spins over time reveals bi-exponential decay curves. A possible reason for this observation would be the presence of two sectors with different spin densities within the mode volume.
This is suggested by several tests. First, we repeat the experiment with the pump far detuned from any spin transition and observe no shift in the resonance frequency. This excludes the possibility that one of the
time constants results from the pump influencing the resonator shift (for example, through heating). Second, we perform the experiment varying the pump pulse frequency around $f_{\mi=+1}$ and observe that the amplitudes of both
exponentials are modulated by the spin linewidth (see Fig.~\ref{T1}). This excludes the possibility
of a broad background spin ensemble such as paramagnetic impurities on the surface of NbTiN or diamond itself. In addition, the ratio of $\gens(T=0)$ values for the two modes differs from the expected $\sqrt{3}$ by a factor $\sim 1.6$. This further hints at a strong inhomogeneity of P1 concentration across the diamond.

\section*{PUMPING AND DEPOLARIZATION OF ENSEMBLES}
\begin{figure}[t]
\includegraphics[width=\columnwidth]{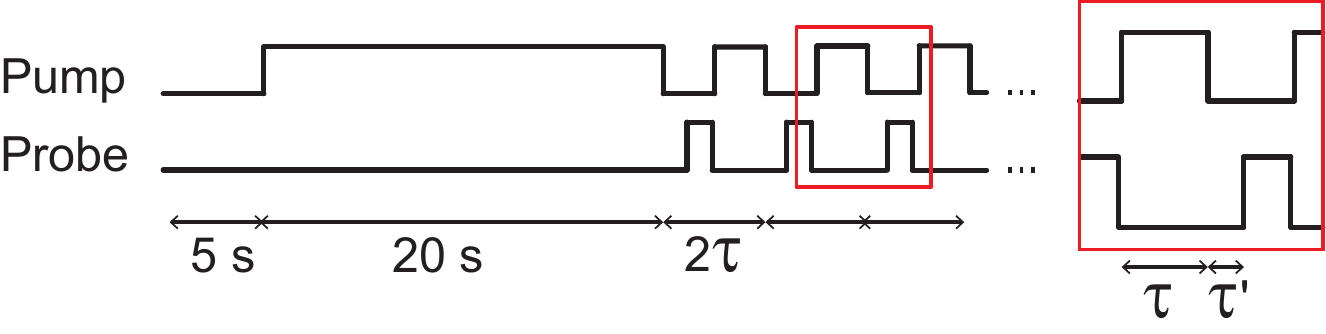}
\caption{Schematic of pump- and probe-tone timings used in Fig.~5. At each $\Bp$ setting, following a $5~\s$ waiting time, pump pulses resonant on driven transitions are applied for $20~\s$. This is followed by a periodic series of pumping and probing steps (period $2\tau$), with the probe frequency $f$ stepped at the start of every cycle. The probe is always on, but $|\Sto|(f)$ only measured for a time $\tau-\tau'$. Here $\tau'$ is a settling time of $20~\millis$. We use $\tau = 100~\millis$ for 2-D transmission spectroscopy [Fig.~5(a)], and $\tau=270~\millis$ for measurements of Rabi-split dip separation as a function of pump power at fixed $\Bp$ [Fig.~5(b)].}
\label{VA2}
\end{figure}

To study the dynamics of spins within and between the sub-ensembles, we resonantly pump the $\mi=0$ transition and measure the effect on the sub-ensembles. The pulse sequence used for transmission spectroscopy $|\Sto|(f,\Bp)$ in Fig.~5(a) is shown in Fig.~\ref{VA2}.

We model the depolarization of spin ensembles due to pumping with a simple rate equation. In addition to the pumping rate $\Omegam$ for the driven transition $(\mi = 0)$, we include a spin-equilibration rate $\Gammao$ (spin diffusion rate across the resonator mode volume) within each sub-ensemble and a nominal rate $\Gamma$ that characterizes cross-relaxation between the sub-ensembles. The rate of change of polarization (equal for $\mi =\pm1$ due to symmetry), is modeled by

\begin{equation}
\frac{d}{dt} \bar P= -M \bar P+ \Gamma
\label{rate},
\end{equation}
with
\[
\bar P=\begin{bmatrix}
			  \bar P(\pm1)\\
  			  \bar P(0)\\
\end{bmatrix},
\]
\[
\Gamma =\Gammao\begin{bmatrix}
			  1\\
			  1\\
\end{bmatrix},
\]
and
\[M =
\begin{bmatrix}
				  \Gammao + \Gamma &  -\Gamma\\
				  -2\Gamma & \Gammao+\Omegam+2\Gamma\\
\end{bmatrix}.
\]
The steady-state polarization ($d\bar P/dt=0$ condition) is $\bar P = M^{-1} \Gamma$, yielding
\[
\bar P(\pm1) = \frac {\Gammao(\Gammao+3\Gamma+\Omegam)}{\Gammao(\Gammao+3\Gamma+\Omegam)+\Gamma\Omegam}.
\]
For $\Gamma\gg\Gammao$~\cite{Sorokin1960}, this expression simplifies to $\bar P(\pm1) = \bar P(0) $ = $\Gammao/(\Gammao+\Omegam/3)$. In this limit, the fast equilibration between sub-ensembles causes pumping of each transition with an effective rate $\Omega/3$, even though only one transition is directly driven by the pump.

